\documentclass[reprint, superscriptaddress, nofootinbib, amsmath,amssymb, aps, prl]{revtex4-1}

\usepackage{graphicx}	
\usepackage{dcolumn}	
\usepackage{bm}
\usepackage[colorlinks]{hyperref}
\usepackage{color}
\usepackage[dvipsnames]{xcolor}
\usepackage{mathtools}

\definecolor{CiteColor}{rgb}{0,0.5,0}
\hypersetup{citecolor=CiteColor}
\definecolor{RefColor}{rgb}{0.55,0,0}
\hypersetup{linkcolor=RefColor}
\definecolor{darkgreen}{rgb}{0.2,0.7,0.2}
\definecolor{cyan}{rgb}{0,0.9,0.9}
\definecolor{orange}{rgb}{0.9,0.5,0}
\definecolor{magenta}{rgb}{1,0,1}
\definecolor{purple}{rgb}{0.8,0.4,0.8}
\definecolor{gray}{rgb}{0.8242,0.8242,0.8242}
\definecolor{dodgerblue}{rgb}{0.12, 0.56, 1.0}
\definecolor{forestgreen}{rgb}{0.2, 0.6, 0.2}

\begin{document}

\title{Rapid generation of fully relativistic extreme-mass-ratio-inspiral\\waveform templates for LISA data analysis}

\author{Alvin J. K. Chua}
\affiliation{Jet Propulsion Laboratory, California Institute of Technology, Pasadena, CA 91109, U.S.A.}

\author{Michael L. Katz}
\affiliation{Department of Physics and Astronomy, Northwestern University, Evanston, IL 60208, United States}
\affiliation{Center for Interdisciplinary Exploration and Research in Astrophysics (CIERA), Evanston, IL, United States}

\author{Niels Warburton}
\affiliation{School of Mathematics and Statistics, University College Dublin, Belfield, Dublin 4, Ireland}

\author{Scott A.~Hughes}
\affiliation{Department of Physics and MIT Kavli Institute, Massachusetts Institute of Technology, Cambridge, MA 02139, USA}

\date{\today}

\begin{abstract}
	The future space mission LISA will observe a wealth of gravitational-wave sources at millihertz frequencies. Of these, the extreme-mass-ratio inspirals of compact objects into massive black holes are the only sources that combine the challenges of strong-field complexity with that of long-lived signals. Such signals are found and characterized by comparing them against a large number of accurate waveform templates during data analysis, but the rapid generation of such templates is hindered by computing the $\sim10^3$--$10^5$ harmonic modes in a fully relativistic waveform. We use order-reduction and deep-learning techniques to derive a global fit for these modes, and implement it in a complete waveform framework with hardware acceleration. Our high-fidelity waveforms can be generated in under $1\,\mathrm{s}$, and achieve a mismatch of $\lesssim5\times 10^{-4}$ against reference waveforms that take $\gtrsim10^4$ times longer. This marks the first time that analysis-length waveforms with full harmonic content can be produced on timescales useful for direct implementation in LISA analysis algorithms.	
\end{abstract}


\maketitle

\paragraph{Introduction.}

As gravitational-wave (GW) astronomy continues to bear fruit \cite{O1-2}, preparatory work is underway for a future generation of ground- and space-based observatories that span the astrophysical GW spectrum \cite{LVK2018LivingReview,PTA,LISAMissionProposal}, and whose success will depend on further advancements in the technology and methods of GW detection. The theorist's contribution to this endeavor lies primarily in the construction of \emph{waveform} models to describe GW signals from astrophysical phenomena, as well as the application of statistical analysis to infer their presence in noisy data and their source properties. For highly relativistic sources, the computational burden of solving Einstein's equations in numerical modeling is fundamentally at odds with the Monte-Carlo nature of modern signal-processing and Bayesian-inference techniques.

Extreme-mass-ratio inspirals (EMRIs) are the most conspicuous example of such dissonance. These are the late capture orbits of stellar-mass ($\mu\sim1$--$100\,M_\odot$) compact objects into the massive ($M\sim10^5$--$10^7\,M_\odot$) black holes in galactic nuclei. They radiate millihertz GWs, and will be a key source class for the space mission LISA \cite{LISAMissionProposal} upon its launch in the next decade. An EMRI signal typically has $\sim10^5$ observable cycles carrying the imprint of the compact object's complex dynamical motion deep in the central black hole's gravitational field. This wealth of information is double-edged: it will allow probes of galactic-nuclei astrophysics and strong-field gravity to unprecedented precision \cite{PhysRevD.95.103012,berry2019unique}, but it places exacting constraints on the accuracy and efficiency of both modeling and data analysis for EMRIs --- to a combined extent far surpassing that for other important LISA sources.

Calculations from black-hole perturbation theory, and in particular from the ongoing gravitational self-force program \cite{2019RPPh...82a6904B}, are on target to produce EMRI waveforms that meet the accuracy requirements of LISA science \cite{PhysRevLett.124.021101,Miller:2020bft}. Such models are computationally intensive, and hence ill suited for direct use in analysis algorithms that are tailored to the EMRI problem \cite{2004CQGra..21S1595G,2009CQGra..26m5004B,2011CQGra..28i4016C,PhysRevD.86.104050,PhysRevD.101.044027}. As in the case of numerical-relativity waveforms for comparable-mass binaries, self-force waveforms must be supplemented and approximated by \emph{template} models that are: i) efficiency-oriented, ii) extensive in their description of both intrinsic and extrinsic effects, and iii) end-to-end from source parameters to detector response. The challenge is to achieve this with a controlled and tolerable loss of accuracy. Strategies developed for the comparable-mass case, such as the standard construction of reduced-order-modeling (ROM) surrogates (e.g., \cite{2019arXiv191010473R}), are less likely to scale feasibly to the signal duration, harmonic complexity and information volume of the full EMRI problem.

The semi-relativistic ``kludges'' \cite{PhysRevD.69.082005,PhysRevD.75.024005,2015CQGra..32w2002C,Chua:2017ujo,EMRIKludgeSuite} are the only existing examples of EMRI template models. Kludges trade accuracy for efficiency by means of a modular build and various computational approximations.  Their common distinguishing feature is a reliance on some weak-field assumption at one or more stages of their construction. The speed and generality of kludge models has greatly facilitated numerous LISA studies on mission performance, data analysis approaches and potential scientific applications. However, kludges incur significant error with respect to fully relativistic models for many sources in the observable space of EMRIs \cite{PhysRevD.75.024005}, and have little room for improvement due to the limitations of the weak-field assumption. This inherent cap on accuracy may count against the continued development and adoption of kludge models, at least in the long term.

In this Letter, we report promising headway against the main obstacle to the rapid generation of fully relativistic EMRI waveforms: efficiently computing the slowly evolving \emph{amplitudes} of the $\sim10^3$--$10^5$ harmonic modes that comprise a single waveform in the canonical angular and frequency-based decomposition \cite{PhysRevD.61.084004,PhysRevD.73.024027}. Through the integration of ROM and deep-learning techniques \cite{PhysRevLett.122.211101}, an analytic model for these amplitudes is fitted to numerical data from a frequency-domain Teukolsky solver \cite{BHPToolkit}. The key to our approach is the use of regression rather than interpolation, resulting in a less precise but global fit that returns the full set of amplitudes simultaneously. This allows the inclusion of relativistic amplitudes in template models, where they are combined with existing fast methods for generating the phasing \emph{trajectories} to varying levels of accuracy.

As the mode-amplitude model is a neural network, it is composed of simple linear-algebra operations and hence amenable to acceleration through a highly parallelized implementation for graphics processing units (GPUs). We exploit this to construct the first EMRI waveform model with sub-second runtimes in a realistic setting, i.e., analysis-length signals ($\sim10^7M$ at sampling rate $1/(2M)$), and full harmonic content (retaining up to $1-10^{-9}$ of total power at initial orbital eccentricities of up to $0.7$). The present model describes the source-frame GW field for eccentric orbits in Schwarzschild, with inspiral trajectories that are accurate at adiabatic order. Our code infrastructure is designed with the end goal of providing analysis-ready template models; specifically, it will readily accommodate post-adiabatic trajectories informed by future self-force calculations, as well as the eventual extension to generic Kerr orbits and the integration of a compatible LISA response model.

\paragraph{Adiabatic waveforms.}
An EMRI's disparate masses $(M,\mu\ll M)$ create a wide separation between its orbital and radiation-reaction timescales. This allows EMRIs to be modelled through a two-timescale expansion \cite{Hinderer:2008dm}. In the leading adiabatic part of this expansion, the equations of motion follow from flux balance laws. Though a purely adiabatic treatment of waveform phasing will be insufficiently accurate to describe a typical EMRI signal over its full duration \cite{Drasco:2005is}, adiabatic trajectories can still be used for data analysis within a hierarchical semi-coherent \emph{search} scheme \cite{2004CQGra..21S1595G,Chua:2017ujo}, or for more slowly evolving binaries with $\mu/M<10^{-6}$ \cite{Gourgoulhon:2019iyu,Amaro-Seoane:2019umn}. Some of the most important post-adiabatic phase effects can be easily added by allowing a particular phase to evolve on the long timescale \cite{Hughes+_inprep}. The computation of mode amplitudes is also only required at adiabatic order \cite{Miller:2020bft}, even for the most stringent analysis task of \emph{inference}. This is due to the disproportionate dependence of GW matched filtering on waveform phasing, rather than its amplitude.

For an EMRI with a non-rotating central black hole, the adiabatic evolution of the orbital energy $\mathcal{E}$ and angular momentum $\mathcal{L}$ is given by $(\dot{\mathcal{E}},\dot{\mathcal{L}})=-(\dot{E},\dot{L})$, where an overdot denotes differentiation with respect to coordinate time $t$, and $(\dot{E},\dot{L})$ is the total flux of energy and angular momentum radiated through null infinity and the event horizon. It is useful to parametrize the system by an equivalent set of quasi-Keplerian orbital elements: the semi-latus rectum (henceforth ``separation'') $p$ and eccentricity $e$, with $\mathcal{E}^2 = p^{-1}(p-2-2e)(p-2+2e)/(p-3-e^2)$ and $\mathcal{L}^2= p^2 M^2/(p-3-e^2)$ \cite{Cutler:1994pb}.
In this parametrization, stable bound orbits exist for $p > p_s = 6+2e$ and $0\le e <1$, where $p_s$ denotes the separatrix \cite{Stein:2019buj}. Each instantaneous orbit $(p,e)$ is associated with a radial and azimuthal frequency, denoted by $\Omega_r$ and $\Omega_\varphi$ respectively.

In the Newman--Penrose formalism \cite{doi:10.1063/1.1724257}, the GW field $h$ at null infinity is related to the Weyl curvature scalar $\psi_4$ via $\ddot{h}=2\psi_4$, where $h=h_+ - i h_\times$ with the usual transverse traceless polarizations $h_{+,\times}$. For each orbit $(p,e)$, $\psi_4$ may be obtained by solving the Teukolsky equation \cite{1973ApJ...185..635T} in the frequency domain; this requires a decomposition of the form $\psi_4 = \sum_{lmn} R_{lmn}(r) Y_{lm}(\theta,\varphi) e^{-i \omega_{mn} t}$, where $Y_{lm}$ are spherical harmonics with spin weight $-2$, and $\omega_{mn} = m\Omega_\varphi + n \Omega_r$ are the mode frequencies. 
It is convenient to define and solve for the complex Teukolsky amplitudes $Z^{\infty,H}_{lmn}$, which describe the limiting behaviour of $R_{lmn}$ as $r\to\infty$ and $r\to2M$ respectively \cite{PhysRevD.61.084004}. 

The GW strain for a detector at some suitably distant coordinates $(t,r,\theta,\varphi)$ is then given by \cite{PhysRevD.73.024027}
\begin{align}\label{eq:waveform_decomp}
	h = \frac{1}{r}\sum_{lmn} A_{lmn}(t-r) Y_{lm}(\theta,\varphi) e^{-i\phi_{mn}(t-r)},
\end{align}
where $A_{lmn} = -2 Z^{\infty}_{lmn}/\omega_{mn}^2$, and $ \phi_{mn} =  m \Phi_\varphi + n \Phi_r $ with $\Phi_{r,\varphi}(t) = \int_0^t d\tau\,\Omega_{r,\varphi}(p(\tau),e(\tau))$. In the results we show, we put $t = \phi = 0$ at periastron; other initial conditions are easily accommodated by adjusting the phase of $A_{lmn}$ \cite{Drasco:2005is,Hughes+_inprep}. The sum over modes spans the indices $2\leq l\leq l_\mathrm{max}$, $|m|\leq l$ and $|n|\leq n_\mathrm{max}$, with $l_\mathrm{max}$ and $n_\mathrm{max}$ determined by some convergence criterion (e.g., \cite{10.1143/PTP.121.843}). For the present work, we set $(l_\mathrm{max},n_\mathrm{max})=(10,30)$, resulting in the sum of 7137 modes (but the explicit evaluation of only 3843, by exploiting mode symmetry \cite{PhysRevD.61.084004}).


\paragraph{Fast trajectories.}

To generate fast inspiral trajectories $(p(t),e(t),\Phi_{r,\varphi}(t))$ for use in template models, we need to rapidly evaluate $(\dot{p}, \dot{e})$ across the domain of $(p,e)$. In a \emph{flux-driven} trajectory for adiabatic waveforms, $(\dot{p}, \dot{e})$ is given in terms of the flux $(\dot{E},\dot{L})$ through null infinity and the horizon, which can be calculated directly from the Teukolsky amplitudes \cite{PhysRevD.61.084004}. However, numerical solutions for the amplitudes are computationally costly and can only be precomputed at a limited number of points in $(p,e)$ space. Fast flux-driven trajectories must thus rely on an accurate and efficient interpolation scheme for the fluxes derived from this numerical data.

In this work, we first introduce a new parameter $u = \ln{(p - p_s + 3.9)}$, then calculate Teukolsky amplitudes and fluxes on a uniform grid in $(u,e)$, where $1.37\leq u\leq3.82$ with spacing $0.05$ and $0\leq e\leq0.8$ with spacing $0.025$. The grid in $u$ gives $p\in[p_s+0.03,p_s + 41.6]$ and places more points near the separatrix, where the data varies more rapidly. Before interpolating the flux data, we factor out the leading post-Newtonian (PN) behaviour $(\dot{E}_\mathrm{PN},\dot{L}_\mathrm{PN})$ \cite{Munna:2020juq} to reduce the impact of interpolation error.
We then create bicubic splines for $(\dot{E}/\dot{E}_\mathrm{PN},\dot{L}/\dot{L}_\mathrm{PN})$ over $(u,e)$, with PN factors restored after evaluating the splines. The inspiral trajectory is computed at runtime for initial values $(p_0,e_0)$, by numerically integrating (for $p>p_s+0.1$) the coupled ordinary differential equations $\{\dot{p}, \dot{e}, \dot{\Phi}_{r,\varphi}\}$ with an adaptive eighth-order Runge--Kutta method. As the flux varies on the radiation-reaction timescale $M^2/\mu$, the solution is very smooth.  This permits large integration steps, so generating each trajectory typically takes only a few milliseconds.

Going beyond flux-driven trajectories to make post-adiabatic waveforms requires the inclusion of gravitational self-force corrections \cite{2019RPPh...82a6904B}. This introduces orbital-timescale variations into the equations of motion, which slows the calculation of a \emph{self-forced} trajectory to minutes or even hours \cite{Osburn:2015duj}. Recently, this barrier was overcome using near-identity transformations \cite{vandeMeent:2018rms}, allowing the transformed equations of motion to be evaluated in milliseconds. Key post-adiabatic corrections at second order in the mass ratio \cite{PhysRevLett.124.021101} are being calculated in the two-timescale framework \cite{Miller:2020bft}, which will incorporate a similar averaging procedure. Thus, the generation of the inspiral trajectory is unlikely to constitute a computational bottleneck for post-adiabatic models either.

\paragraph{Neural-network amplitudes.}

With the inspiral trajectory on hand, the remaining computationally nontrivial operation in Eq.~\eqref{eq:waveform_decomp} (besides the sum over modes at high resolution in time) is the evaluation of the mode amplitudes $A_{lmn}(t)$.\footnote{We fit $A_{lmn}$ directly rather than $Z^\infty_{lmn}$, to avoid numerical divergences due to fitting error whenever $\omega_{mn}$ approaches zero.} Although these are very slowly evolving and can be downsampled significantly in time, a conventional spline-interpolation approach requires the creation and evaluation of $>3000$ splines over the $(p,e)$ space. Furthermore, future waveforms for generic Kerr orbits would involve $\sim10^5$ splines over the four-dimensional space of separation, eccentricity, orbital inclination and primary spin. This is problematic, as the ability of most interpolation schemes to simultaneously maintain accuracy and efficiency rapidly degrades for $\gtrsim3$ variables.

To address the issue of high dimensionality (in both the space of modes and the space of orbits), we propose the approach of precomputing an analytic global fit for the mode amplitudes. The particular method we use is \texttt{Roman} \cite{PhysRevLett.122.211101}, which combines the compressive power of ROM with the high-dimensional regression capabilities of deep neural networks. \texttt{Roman} was developed within the paradigm of ROM in GW modeling and analysis \cite{PhysRevLett.106.221102}, and provides an alternative to the combination of surrogate waveforms \cite{PhysRevX.4.031006} with the inference technique of reduced-order quadrature \cite{PhysRevD.87.124005} (albeit at the expense of a more difficult initial fit). However, one open problem with the direct usage of ROM to fit full waveforms is accuracy. While the errors incurred by leading models (e.g., \cite{PhysRevD.96.024058}) are sufficiently small for present ground-based applications, waveform templates for LISA data analysis will require far more stringent modeling \cite{PhysRevD.76.104018}.

In this work, we apply \texttt{Roman} to the fitting of mode amplitudes instead. A greedy algorithm \cite{RomPy} is first used to construct a reduced basis $B$ for (the span of) the Teukolsky amplitude data on the uniform grid in $(u,e)$. This allows the vectorized amplitudes $A_i=\mathrm{vec}(A_{lmn})\in\mathbb{C}^{3843}\cong\mathbb{R}^{7686}$ to be represented in the reduced form
\begin{equation}
A_i(u,e)=\sum_j\alpha_j(u,e)B_{ji}\equiv\alpha_j(u,e),
\end{equation}
where $\alpha_j\in\mathbb{C}^{99}\cong\mathbb{R}^{198}$ for an effective compression factor of around 40. A deep neural network is then trained on the reduced data set $\{u,e,\alpha_\mathrm{num}\}$ as a regression model for $\alpha(u,e)$. The architecture and training of the network is identical to the main example in \cite{PhysRevLett.122.211101}, with the following exceptions: i) Our network contains 20 hidden layers $a_\ell$, where the first six comprise $2^{\ell+1}$ nodes and the remaining layers have 256 nodes each. ii) As the size of the training set is small ($<2000$), Monte Carlo validation \cite{kuhn2013applied} is used to prevent overfitting, with 20 random examples held out at each epoch. iii) The mini-batch size is 810. iv) The loss function is the standard $L^2$ loss $|\alpha-\alpha_\mathrm{num}|^2$ averaged over each mini-batch, where $|\cdot|$ is Hermitian. Our network is trained over $3\times10^4$ epochs, after which it is evaluated at runtime for a set of input points $\{(p,e)\}$ to simultaneously output the corresponding set of mode amplitudes $\{\alpha\cdot B\}$. Finally, we renormalize each amplitude vector by a more accurate estimate for the vector norm, which is directly interpolated using a bicubic spline.

\paragraph{Parallelized implementation.}

The long-lived nature of EMRI signals will necessitate parallel implementations of template models and analysis algorithms, which can then be fully capitalized on through hardware acceleration. To date, accelerator hardware such as GPUs are very underutilized in GW astronomy. However, there are a few examples of GPU usage for both modeling and analysis: the generation of EMRI waveforms with time-domain Teukolsky solvers \cite{2010CoPhC.181.1605K,10.1145/2335755.2335808}; binary-black-hole waveform modeling and population inference for ground-based observing \cite{PhysRevD.100.043030}; as well as massive-black-hole-binary waveform creation and parameter estimation for LISA \cite{PhysRevD.102.023033}.

In this work, our flux-driven trajectory and \texttt{Roman} amplitudes are combined in Eq.~\eqref{eq:waveform_decomp} to form an efficient adiabatic waveform model $h_{+,\times}(t)$ for eccentric Schwarzchild orbits, parametrized by the set $\{M,\mu,p_0,e_0,r,\theta,\varphi\}$. This model is implemented natively for GPUs, with an otherwise-equivalent counterpart implementation for CPUs. The source code is written in \texttt{Python} (interface), \texttt{C++} and \texttt{CUDA}, and is publicly available online \cite{FEW_code}.

GPU acceleration is crucial for relieving the main computational bottleneck in the construction of a time-domain EMRI waveform: the combination and summation of amplitude and phase information at a sufficiently high sampling rate for fully coherent analysis (defined here as $1/(2M)$ for concreteness). In our model, this bottleneck is dealt with through a large-scale cubic-spline interpolation of $A_{lmn}(t)$ and $\Phi_{r,\varphi}(t)$ at a sparse ($\sim10^2$) set of points in time. The number of considered modes is first reduced significantly (to $\sim10^2$--$10^3$) by a runtime selection routine, where all modes at each point in time are sorted by power and removed if they do not contribute cumulatively up to some specified fraction of their total power (typically $\gtrsim1-10^{-5}$ for satisfactory waveform accuracy).
Specific sets of modes can also be chosen for particular analysis purposes, e.g., $l_\mathrm{max}=2$ to search for EMRIs at large separation. The selected amplitude (and phase) splines are then fed into a summation kernel, where they are evaluated and summed at full resolution.

\paragraph{Results.}

\begin{figure}[t]
	\includegraphics[width=9cm]{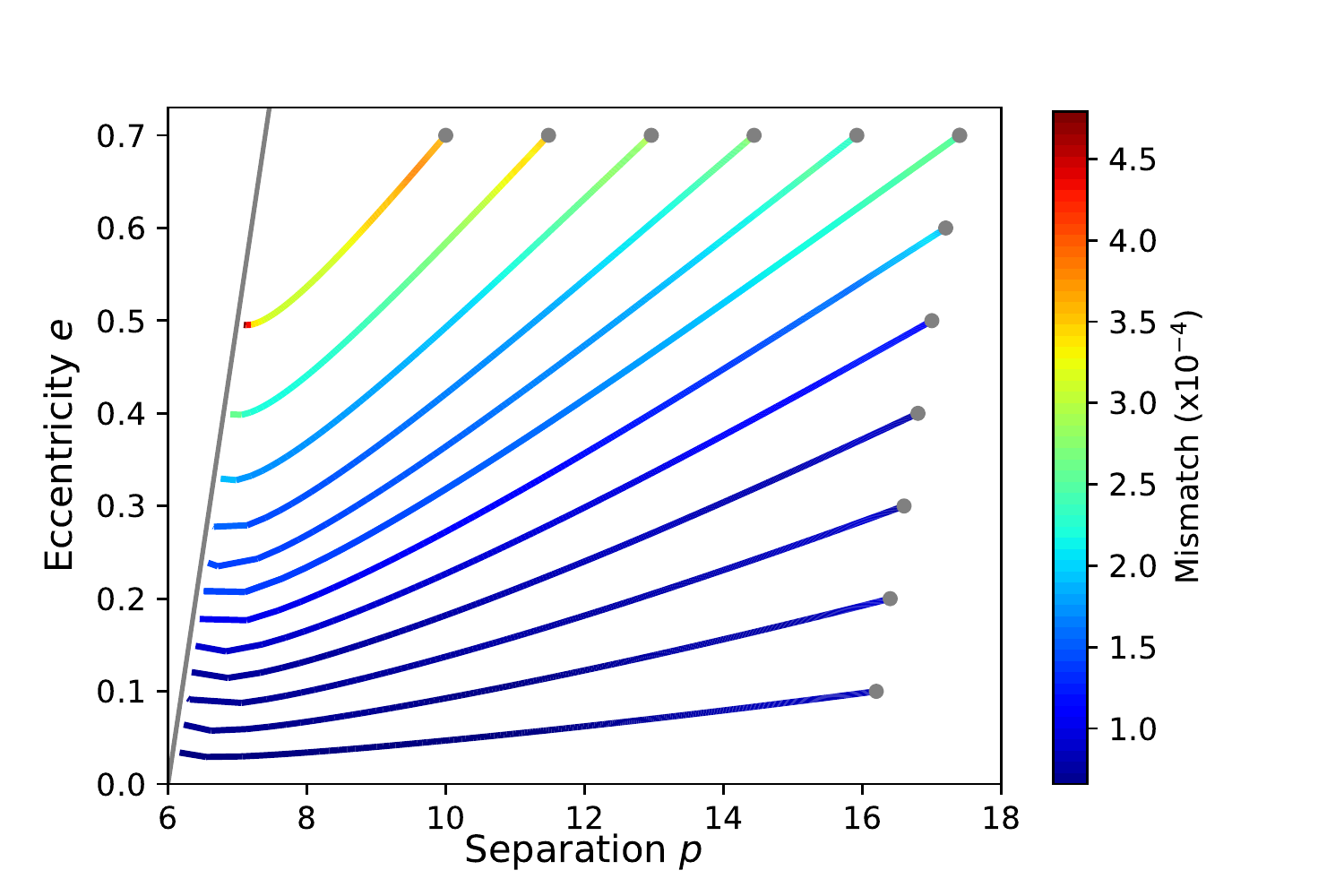}
	\caption{Evolution of mismatch between fast and fiducial waveforms from $(p_0,e_0)$ to $(p,e)$, for 12 EMRIs with $M=10^6M_\odot$, 
$\mu\in[15,304]M_\odot$, and $(p_0,e_0)$ along the model domain boundary. Each small mass is chosen such that the EMRI plunges after a year. These results are for $(\theta,\varphi)=(\pi/2,0)$, but do not depend strongly on the viewing angle. In the worst case (top-left curve), the final $0.01\%$ of the waveform causes the mismatch to increase from under $4\times10^{-4}$ to $5\times10^{-4}$.}
	\label{fig:mismatch}
\end{figure}

The domain of validity for our waveform model is defined as $p_\mathrm{min}\leq p\leq p_s+10$ and $0\leq e\leq0.7$, where $p_\mathrm{min}=\max{\{p_s+0.1,7p_s-41.9\}}$. Orbits at small $p$ and large $e$ are excluded as they lack astrophysical relevance, and are also difficult to fit due to their high degree of variability. The large-$p$ boundary is justified by the reduced sensitivity of LISA at frequencies corresponding to $p\gtrsim20$. We assess the individual accuracies of the trajectory and amplitude modules against numerical Teukolsky flux and amplitude calculations, using a test data set of 232 orbits that spans the domain of validity (but has no orbit in common with the training set). For the relative flux error $(\Delta\dot{E}/\dot{E}_\mathrm{num},\Delta\dot{L}/\dot{L}_\mathrm{num})$, both components have a median value of $3\times10^{-7}$.
As the vectorized \texttt{Roman} amplitudes are renormalized to similar accuracy, we consider their ``mode-distribution'' error $1-\Re{(A^\dagger A_\mathrm{num})}/(|A||A_\mathrm{num}|)$, which reduces to (half of) the relative $L^2$ error when $|A|=|A_\mathrm{num}|$. The mode-distribution error has a median value of $4\times10^{-5}$.

Our fast model is then benchmarked against a slower fiducial model that uses standard bicubic-spline interpolation for the amplitude of each mode (without mode selection), as well as integration steps at full time resolution for the inspiral trajectory. The bicubic amplitudes in the slow model are significantly more faithful to the numerical test data than the \texttt{Roman} amplitudes, with a median mode-distribution error of $3\times10^{-11}$. To quantify the overall error in the fast waveform with respect to the slow fiducial waveform, we examine their mismatch: $1-\Re{(h^\dagger h_\mathrm{fid})}/(|h||h_\mathrm{fid}|)$, defined here without noise weighting for simplicity. The mismatch is dominated by amplitude error, as the phase difference $\Delta\phi$ between the fast and slow trajectories typically has a maximal value of $\sim10^{-3}$ over the full duration of a waveform.

\begin{figure}[t]
	\includegraphics[width=9cm]{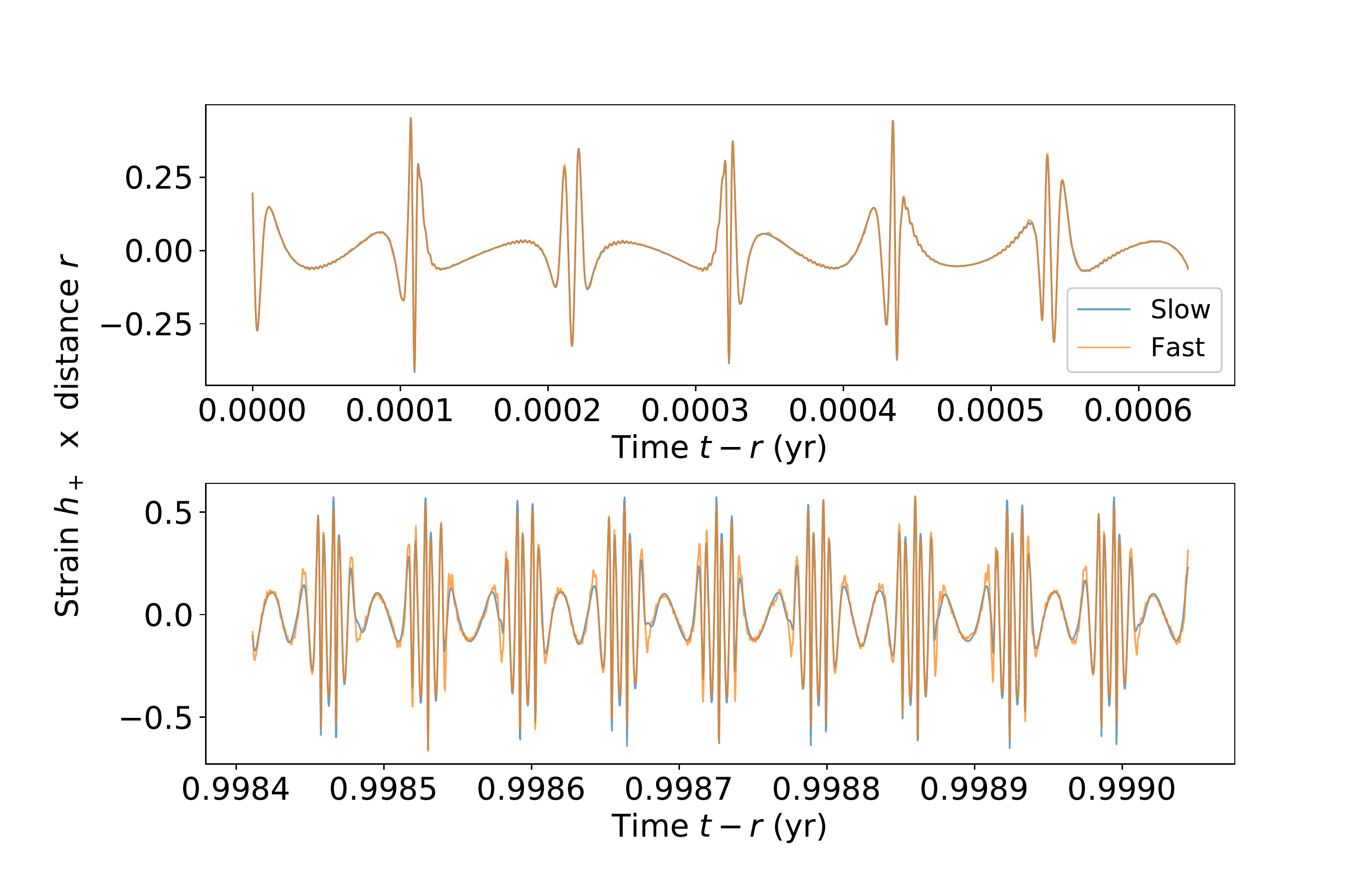}
	\caption{Six-hour snapshots of fast (orange) and fiducial (blue) waveforms, one year before plunge (top) and just before plunge (bottom). Waveforms are for the worst-case EMRI $(M,\mu,p_0,e_0)=(10^{6}M_\odot, 15 M_\odot, 10, 0.7)$, with a one-year mismatch of $5\times 10^{-4}$. Small amplitude deviations are visible just before plunge at $(p,e)\approx(7,0.5)$, where the mode-distribution error approaches its maximum across the domain of validity.}
	\label{fig:waveform}
\end{figure}

\begin{figure}[t]
	\includegraphics[width=9cm]{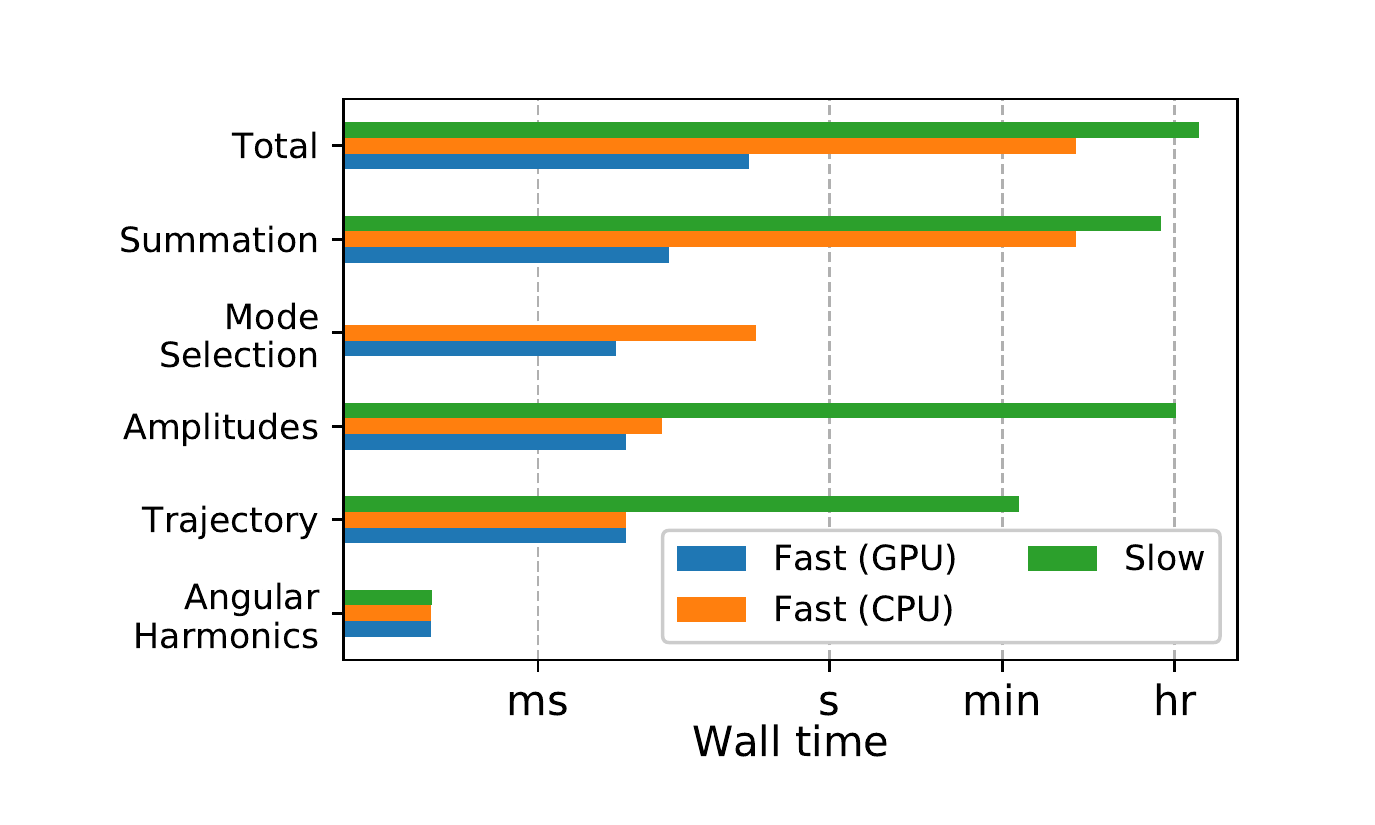}
	\caption{Computational wall time for fast and fiducial waveforms, broken down into individual modules. All times are averaged over $\geq5$ evaluations of the worst-case waveform on a single CPU core (and GPU), where the CPU is an Intel Xeon Gold 6132 and the GPU is an NVIDIA Tesla V100.}
	\label{fig:timing}
\end{figure}

Fig.~\ref{fig:mismatch} shows how the mismatch from $(p_0,e_0)$ up to $(p,e)$ changes for a representative set of EMRIs in the domain of validity, as they evolve towards the separatrix over a duration of $\sim10^7M$. The EMRI with the largest $e_0$ and the smallest $p_0$ plunges at high eccentricity $e\approx0.5$, and has the ``worst-case'' full mismatch of $5\times10^{-4}$; snapshots of this waveform at initial and plunge time are shown in Fig.~\ref{fig:waveform}. In general, the signal-to-noise ratio $\rho$ of a GW source determines the required level of mismatch $\sim1/\rho^2$ for inference purposes  \cite{PhysRevD.76.104018}, and so the accuracies achieved by a waveform with \texttt{Roman} amplitudes should be adequate for LISA EMRIs (where $\rho\lesssim10^2$). In terms of efficiency, wall times for the slow model ($\sim1\,\mathrm{hr}$ for the worst-case waveform with $\sim10^3$ modes) are dominated not just by mode summation, but also the evaluation of mode amplitudes (see Fig.~\ref{fig:timing}). This is not the case for our fast model, where the bottleneck is reduced solely to summation, and wall times are reduced to $\sim1\,\mathrm{min}$ on a CPU and further to $\sim10^2\,\mathrm{ms}$ on a GPU.

\paragraph{Conclusion.}

The efficient computation of fully relativistic EMRI waveform templates has yet to be achieved under the constraints of LISA data analysis, as a significant bottleneck is posed by the interpolation and evaluation of the $\sim10^3$--$10^5$ mode amplitudes. In this Letter, we propose that the bottleneck can first be relieved by combining order-reduction and deep-learning techniques in the amplitude fit \cite{PhysRevLett.122.211101}, and then virtually removed through the use of GPU acceleration. We demonstrate this by introducing the first EMRI waveform model with sub-second runtimes for analysis-length signals with full harmonic content. Access to higher modes during analysis is important not just for precise inference, but also for finding signals in the first place: using our model, we find that a quadrupolar waveform with $l_\mathrm{max}=2$ typically has a mismatch of $\approx0.1$ against a fiducial waveform, which may be suboptimal even for search \cite{Chua:2017ujo}.

Our present waveform model is accurate at adiabatic order for eccentric Schwarzschild orbits, and thus can already be used to construct search templates for EMRIs with a non-rotating large mass. However, LISA data analysis needs template models that describe generic Kerr EMRIs at sufficient accuracy for inference. The framework presented in this Letter is designed to accommodate the increased accuracy and extensiveness of such models while retaining efficiency. Post-adiabatic waveforms require the replacement of flux-driven trajectories with self-forced trajectories, which will be equally efficient \cite{vandeMeent:2018rms,Miller:2020bft}. Practical schemes for dealing with transient resonances \cite{PhysRevD.83.044037,PhysRevD.89.084028,PhysRevLett.114.081102} can be included as well. Although the mode amplitudes are required only at leading order \cite{Miller:2020bft}, they must be extended to cover the space of Kerr orbits; our fitting technique is promising for dealing with the increased dimensionality.
The source-end waveform also has to be integrated with a realistic LISA response, which could be done through a frequency-domain approximation for both waveform \cite{Hughes+_inprep} and response \cite{2009CQGra..26m5004B,PhysRevD.101.044027}, or by developing accelerated versions of more accurate time-domain simulators \cite{PhysRevD.71.022001,PhysRevD.77.023002}. Finally, the modular nature of the framework allows the incorporation of additional physics as well. This could include environmental effects, e.g., accretion disks \cite{Kocsis:2011dr} and massive perturbers \cite{Yang:2017aht,Bonga:2019ycj}, or new physics, e.g., beyond-general-relativity corrections \cite{Maselli:2020zgv} and beyond-standard-model physics \cite{Hannuksela:2018izj, Hannuksela:2019vip}.

\paragraph{Acknowledgements.}
AJKC acknowledges support from the Jet Propulsion Laboratory (JPL) Research and Technology Development program, and from the NASA grant 18-LPS18-0027. MLK acknowledges support from the National Science Foundation under grant DGE-0948017. NW gratefully acknowledges support from a Royal Society -- Science Foundation Ireland University Research Fellowship. SAH is supported by NASA ATP Grant 80NSSC18K1091 and NSF Grant PHY-1707549. We thank Chad Galley for the RomPy package \cite{RomPy} and discussions on reduced-order modeling, as well as Ian Hinder and Barry Wardell for the SimulationTools analysis package \cite{SimulationTools}. Final thanks go out to Soichiro Isoyama and Adam Pound for helpful discussions. This work makes use of the Black Hole Perturbation Toolkit \cite{BHPToolkit}. Our research was supported in part through the computational resources and staff contributions provided for the Quest/Grail high performance computing facility at Northwestern University. Parts of this work were carried out at JPL, California Institute of Technology, under a contract with the National Aeronautics and Space Administration. \copyright\,2020. All rights reserved.

\nocite{*}

\bibliography{references}

\end{document}